\documentclass[conference]{IEEEtran}
\IEEEoverridecommandlockouts
\usepackage{cite}
\usepackage{amsmath,amssymb,amsfonts}
\usepackage{algorithmic}
\usepackage{graphicx}
\usepackage{textcomp}
\usepackage{xcolor}
\usepackage{mathrsfs}
\usepackage{bm}
\usepackage[ruled,linesnumbered]{algorithm2e}
\usepackage[utf8]{inputenc}
\usepackage{float}
\usepackage{url}
\usepackage{flushend}
\usepackage{subfigure}
\usepackage{todonotes}

\def\BibTeX{{\rm B\kern-.05em{\sc i\kern-.025em b}\kern-.08em
    T\kern-.1667em\lower.7ex\hbox{E}\kern-.125emX}}
\begin{document}

\title{Text-Guided Token Communication for Wireless Image Transmission\\}
\author{
	\IEEEauthorblockN{Bole Liu\IEEEauthorrefmark{1}, Li Qiao\IEEEauthorrefmark{1}, Ye Wang\IEEEauthorrefmark{1}, Zhen Gao\IEEEauthorrefmark{2}, Yu 
    Ma\IEEEauthorrefmark{3}, Keke 
    Ying\IEEEauthorrefmark{1}, Tong 
    Qin\IEEEauthorrefmark{1}
    }
    
	\IEEEauthorblockA{\IEEEauthorrefmark{1}School of Information and Electronics, Beijing Institute of Technology, Beijing 100081, China}
    \IEEEauthorblockA{\IEEEauthorrefmark{2}Advanced Research Institute of Multidisciplinary Science, Beijing Institute of Technology, Beijing 100081, China}
    \IEEEauthorblockA{\IEEEauthorrefmark{3}School of Future Technologies, Beijing Institute of Technology, Beijing 100081, China}
	Email: \url{gaozhen16@bit.edu.cn}}

\maketitle

\begin{abstract}
With the emergence of 6G networks and proliferation of visual applications, efficient image transmission under adverse channel conditions is critical. We present a text-guided token communication system leveraging pre-trained foundation models for wireless image transmission. Our approach converts images to discrete tokens, applies 5G NR polar codec on top of the tokenizeation, and employs text as a conditioning signal to generate lost tokens to mitigate the cliff effect at lower signal-to-noise ratios (SNRs). Evaluations on ImageNet show our method outperforms state-of-the-art deep joint source-channel coding scheme in perceptual quality and semantic preservation at extremely low bandwidth ratio, i.e., 1/96. In addition, Our system requires no scenario-specific retraining and exhibits superior cross-dataset generalization, establishing a new paradigm for efficient image transmission aligned with human perceptual priorities.
\end{abstract}

\section{Introduction}
The emergence of 6G networks promises transformative advancements in wireless communication \cite{chen20236g, gao20246g, Wang20236g}, supporting novel applications that require robust transmission of high-quality visual content under challenging channel conditions. As these applications proliferate at network edges, efficient image transmission under adverse channel conditions has emerged as a critical research challenge \cite{Liu2024space}. Deep learning has enabled deep joint source channel coding (D-JSCC) for semantic communications that optimize transmission pipeline end-to-end \cite{bourtsoulatze2019deep,xu2021wireless, wu2025semantic}. However, current D-JSCC approaches often overlook hardware precision constraints\cite{shi2021semantic}, and require computationally expensive retraining for new environments or channel conditions. These challenges have motivated exploration of alternative approaches leveraging pre-trained generative foundation models and generative semantic communication \cite{qiao2024wcl, gao2023semantic}.


Tokens serve as the currency and fundamental processing units of generative foundation models. In particular, learned vector quantization has emerged as a popular paradigm for image tokenization, evolving from VQVAE \cite{van2017neural} to enhanced frameworks like VQGAN \cite{esser2021taming}. Furthermore, with the development of multimodal large language models, multimodal tokenizers such as text-aware transformer-based 1D Tokenizer (TA-TiTok)\cite{kim2025demo}, Muse\cite{muse}, enable semantic alignment between visual and textual tokens. These multimodal tokenizers preserve both visual details (via image tokens) and semantic concepts (via text tokens), offering unique advantages for communication systems, as discrete tokens naturally align with digital transmission constraints. Bi-directional or uni-directional transformers can be trained on top of image tokenizers to enable masked token prediction and next-token generation. These unique properties of tokens and pre-trained generative foundation models have been leveraged in generative semantic communications in \cite{qiao2025tokcom}, termed token communication (TokCom), which provides semantic resilience against channel impairments. The authors of \cite{lee2025semantic} further demonstrated that optimizing token packetization can enhance TokCom performance. In addition, TokCom has also been shown to benefit next-generation multiple access schemes \cite{qiao2025todma,qiao2025todma_large}.

While TokCom has been attracting increasing attention, it is still in its infancy. Recent studies have not yet addressed how and to what extent text tokens can guide and improve TokCom performance. Moreover, compared to state-of-the-art D-JSCC schemes, it remains unclear whether and in which aspects TokCom can outperform D-JSCC. These questions remain open and require further investigation.

To solve these problems, this paper propose a novel text-guided token communication system for wireless image transmission that leverages pre-trained multi-modal foundation models to achieve reliable communication with extremely low bandwidth ratio, even under adverse channel conditions. Our contributions are twofold:
\begin{itemize}
    \item We design a novel text-guided token communication system utilizing text tokens as auxiliary information for image transmission, demonstrating superior performance at signal-to-noise ratios (SNRs) above 0dB with bandwidth ratio of 1/96 while mitigating the cliff effect at low SNR conditions.
    
    \item We quantitatively validate the effectiveness of text guidance and analyze the importance of different modalities through varying text lengths and token error rates (TERs) in this system. Additionally, we examine the system's generalization capabilities, demonstrating its ability to adapt to diverse channel scenarios and dataset distributions without retraining, requiring only adjustments to channel coding parameters when necessary. 

\end{itemize}

\textit{Notation}: We denote scalars and matrices
by lower-case letters and uppercase boldface letters, respectively. 

\section{System Model}
\begin{figure*}[htbp]
    \centering
    \includegraphics[width=0.9\textwidth]{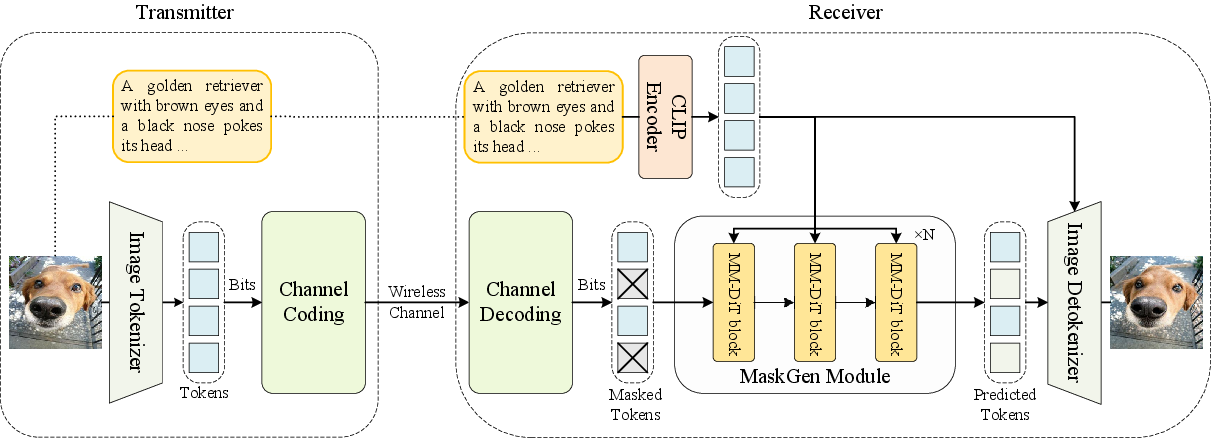}
    \caption{Overall architecture of the proposed text-guided token communication system for wireless image transmission. The pipeline consists of: (1) image tokenization using a pre-trained vector quantized tokenizer; (2) channel coding with 5G NR-compliant polar codes; (3) channel transmission over AWGN with varying SNR levels; and (4) text-guided image reconstruction with masked token prediction.}
    \label{fig:discrete_frame}
    \vspace{-10pt}
\end{figure*}

We propose a novel text-guided token communication system leveraging pre-trained image tokenizers for efficient image transmission. Our system assumes the transmitter and receiver share common textual prior knowledge that serves as semantic guidance for reconstruction. Fig.~\ref{fig:discrete_frame} illustrates our system architecture with four components: image tokenization, channel coding/decoding, transmission, and token-based reconstruction. 

\subsection{Image Tokenization}
The image tokenization module employs a TA-TiTok\cite{kim2025demo} image tokenizer to convert input images into discrete token representations. Unlike traditional image compression approaches that operate at the pixel level, this tokenizer extracts semantic-level features and maps them to indices in a learned codebook.

The tokenization process can be mathematically expressed as
\begin{equation}
\mathbf{T} = \mathcal{E}(\mathbf{X}),
\end{equation}
where $\mathbf{X} \in \mathbb{R}^{H \times W \times 3}$ is the input RGB image resized to 256 × 256 and normalized to the range [0,1], $\mathcal{E}(\cdot)$ is the encoder function, and $\mathbf{T} = [T_1, T_2, \ldots, T_{128}]$ is the resulting sequence of 128 token indices. The learned codebook $\mathbf{\mathcal{C}} \in \mathbb{R}^{N \times D}$ maps each token index $T_i$ to a corresponding embedding vector $\mathbf{\mathcal{C}}_{T_i} \in \mathbb{R}^D$, where $N$ is the codebook size and $D$ is the embedding dimensionality. This discrete representation significantly reduces the data volume required for transmission while preserving essential semantic information.

\subsection{Channel Coding}

After tokenization, the transmission task is reduced to reliably sending token indices. To protect these token indices against channel noise, we employ polar coding following the 5G NR standard (3GPP TS 38.212 \cite{3gpp38212}).

For practical implementation, we partition the 128 tokens into $P$ equal-sized packages. We then apply standard 5G NR coding with a code length of $B$ bits per package, following this pipeline:

\begin{enumerate}
    \item \textbf{Cyclic Redundancy Check (CRC) Attachment}: We append an 11-bit CRC to each information block using the standard 5G NR generator polynomial $g(D) = D^{11} + D^{10} + D^{9} + D^{5} + 1$.

    \item \textbf{Polar Encoding}: The processed bits are encoded using the standard polar code generator matrix.
    
    \item \textbf{Channel Interleaving}: The CRC-attached bits are interleaved using a triangular-based channel interleaver as defined in the 5G NR standard.
    
    \item \textbf{Rate Matching}: Rate matching is applied to ensure the output length matches the target transmission length through bit selection or repetition via a circular buffer.
\end{enumerate}

Our implementation uses the NVIDIA-Sionna library \cite{sionna} with Successive Cancellation List (SCL) decoding with a list size of 8.

\subsection{Channel Transmission}

The encoded symbols are modulated using 4-QAM and transmitted through an additive white gaussian noise (AWGN) channel, with the received signal represented as
\begin{equation}
\mathbf{Y} = \mathbf{X}_m + \mathbf{N},
\end{equation}

where $\mathbf{X}_m$ represents the modulated symbols, and $\mathbf{N} \sim \mathcal{N}(0, \sigma^2)$ represents the channel noise with $\sigma^2$ denoting the noise power.

Akin to \cite{bourtsoulatze2019deep,xu2021wireless,shi2021semantic}, the bandwidth ratio is defined as the ratio of transmitted symbols to the total number of pixels in the original image

\begin{equation}
R = \frac{\text{Symbols}}{H \times W \times c},
\end{equation}
where Symbols represents the number of transmitted modulation symbols, and $H \times W \times c$ denotes the total pixels in an RGB image with height $H$, width $W$ and channel $c$.

\subsection{Token-Based Image Reconstruction}

At the receiver side, we employ a multi-stage decoding process that combines traditional error correction with multimodal foundation model based token restoration.

\subsubsection{Channel Decoding}
First, we perform polar decoding on the received signal using SCL decoding
\begin{equation}
\hat{\mathbf{B}}_{int} = \text{Polar}_{\text{SCL}}(\mathbf{Y}, n, k, L),
\end{equation}
where $\text{Polar}_{\text{SCL}}$ represents the SCL decoding function with code length $n$, information length $k$, and list size $L=8$. The decoded bits are then de-interleaved and CRC-verified to obtain the decoded token indices
\begin{equation}
\hat{\mathbf{T}} = \text{BinaryToIndex}(\hat{\mathbf{B}}_{int}).
\end{equation}

\subsubsection{Masked Token Generation}
For token-level error concealment, we employ a pre-trained text-guided masked generative model (MaskGen) \cite{kim2025demo} that can leverage the contextual relationships between tokens to recover corrupted indices. Our text-guided approach consists of:

\begin{enumerate}
    \item \textbf{Error Detection}: We identify potentially corrupted tokens using reliability information from the decoder
    \begin{equation}
        \mathcal{M} = \{i \mid \text{CRC}_i = \text{failure} \},
    \end{equation}
    where $\mathcal{M}$ is the set of positions with potentially corrupted tokens, $\text{CRC}_i$ indicates CRC check status.
    
    \item \textbf{Token Masking}: We mask the identified corrupted tokens
    \begin{equation}
    \hat{\mathbf{T}}_{\text{masked}}[i] = 
    \begin{cases}
        \hat{\mathbf{T}}[i] & \text{if } i \notin \mathcal{M} \\
        \text{[MASK]} & \text{if } i \in \mathcal{M}
    \end{cases}.
    \end{equation}
    
    \item \textbf{Iterative Token Prediction}: The MaskGen model predicts the masked tokens using an iterative confidence-based approach
    \begin{equation}
    \hat{\mathbf{T}}_{\text{corrected}} = \text{MaskGen}(\hat{\mathbf{T}}_{\text{masked}}, \text{text}),
    \end{equation}  
    where $\text{text}$ is the optional text description providing semantic guidance for restoration.
\end{enumerate}

The MaskGen model leverages both spatial context and semantic information from textual descriptions to reconstruct missing visual content.

\subsubsection{Image Reconstruction}
Finally, the corrected token sequence is used to reconstruct the image
\begin{equation}
\hat{\mathbf{X}}_{\text{text}} = \mathcal{D}_{\text{text}}(\hat{\mathbf{T}}_{\text{corrected}}, \text{text}),
\end{equation}
where $\mathcal{D}$ is the text-guided decoder function of the pre-trained vector quantized (VQ) model that leverages semantic information from the text guidance generated by the CLIP encoder \cite{clip}.

\begin{figure*}[t]
    \centering
    \subfigure[PSNR(↑) vs. SNR]{
    \includegraphics[width=0.32\textwidth, trim=8mm 0mm 10mm 0mm, clip]{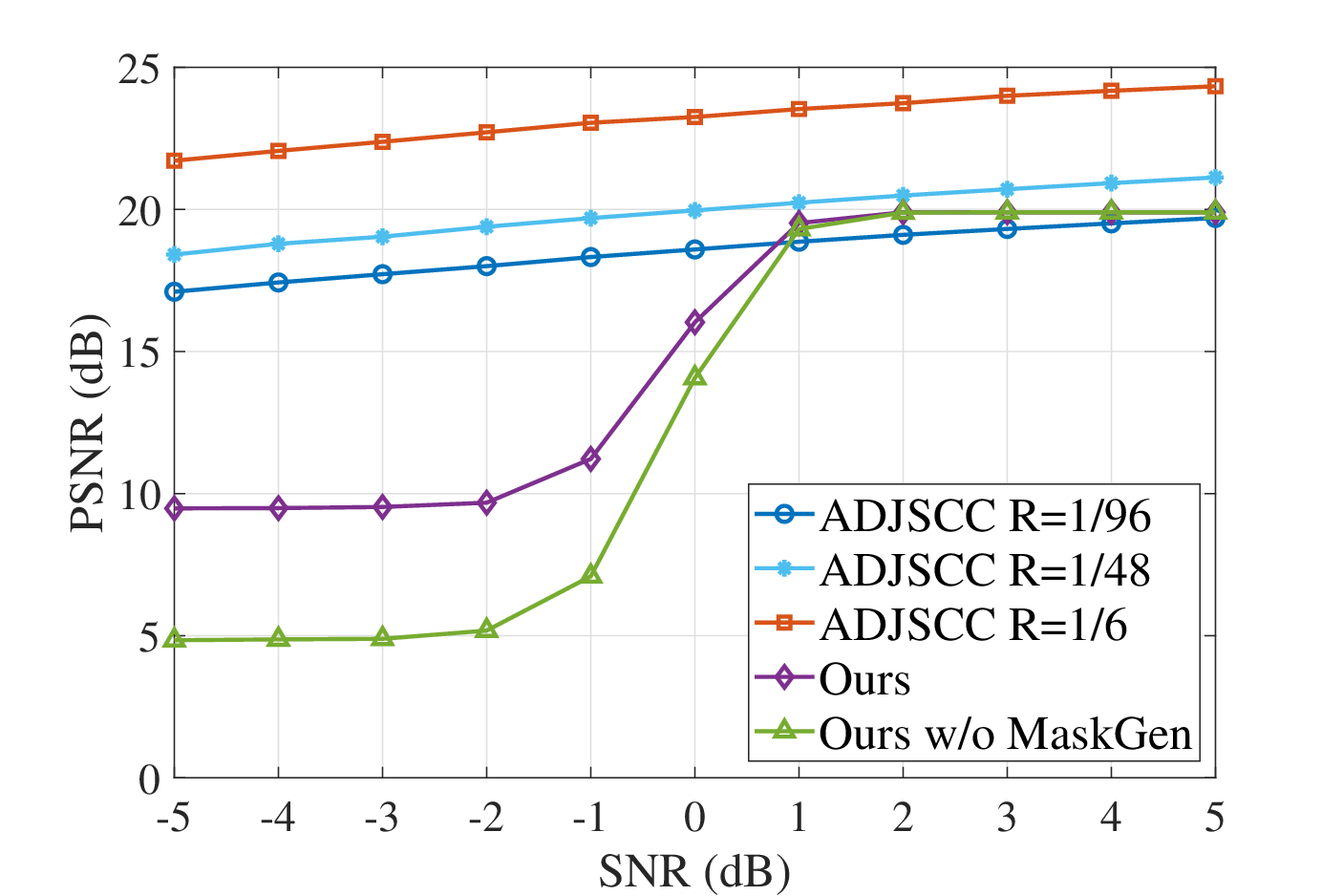}}
    \subfigure[LPIPS(↓) vs. SNR]{
    \includegraphics[width=0.32\textwidth, trim=8mm 0mm 10mm 0mm, clip]{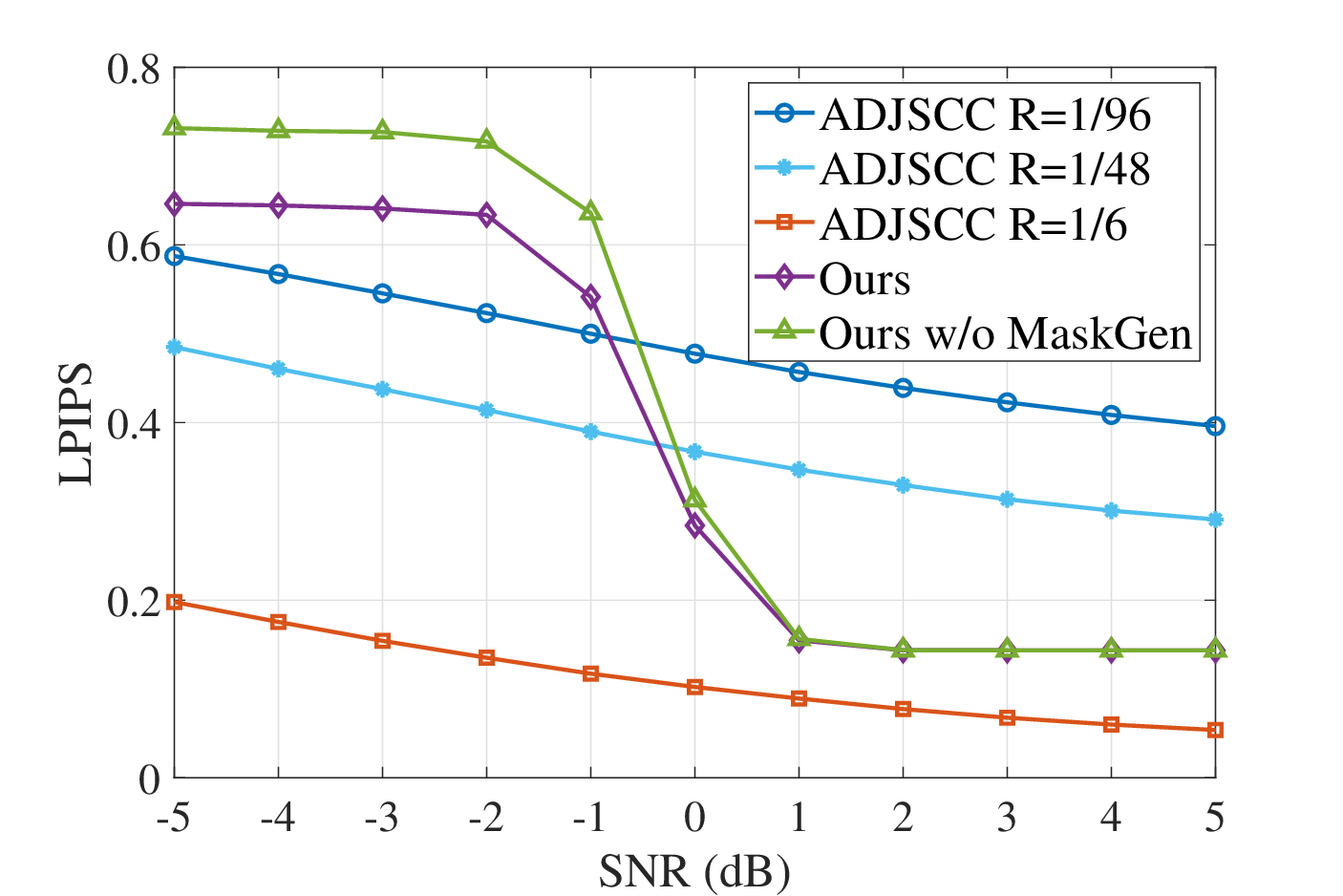}}
    \subfigure[CLIP(↑) vs. SNR]{
    \includegraphics[width=0.32\textwidth, trim=8mm 0mm 10mm 0mm, clip]{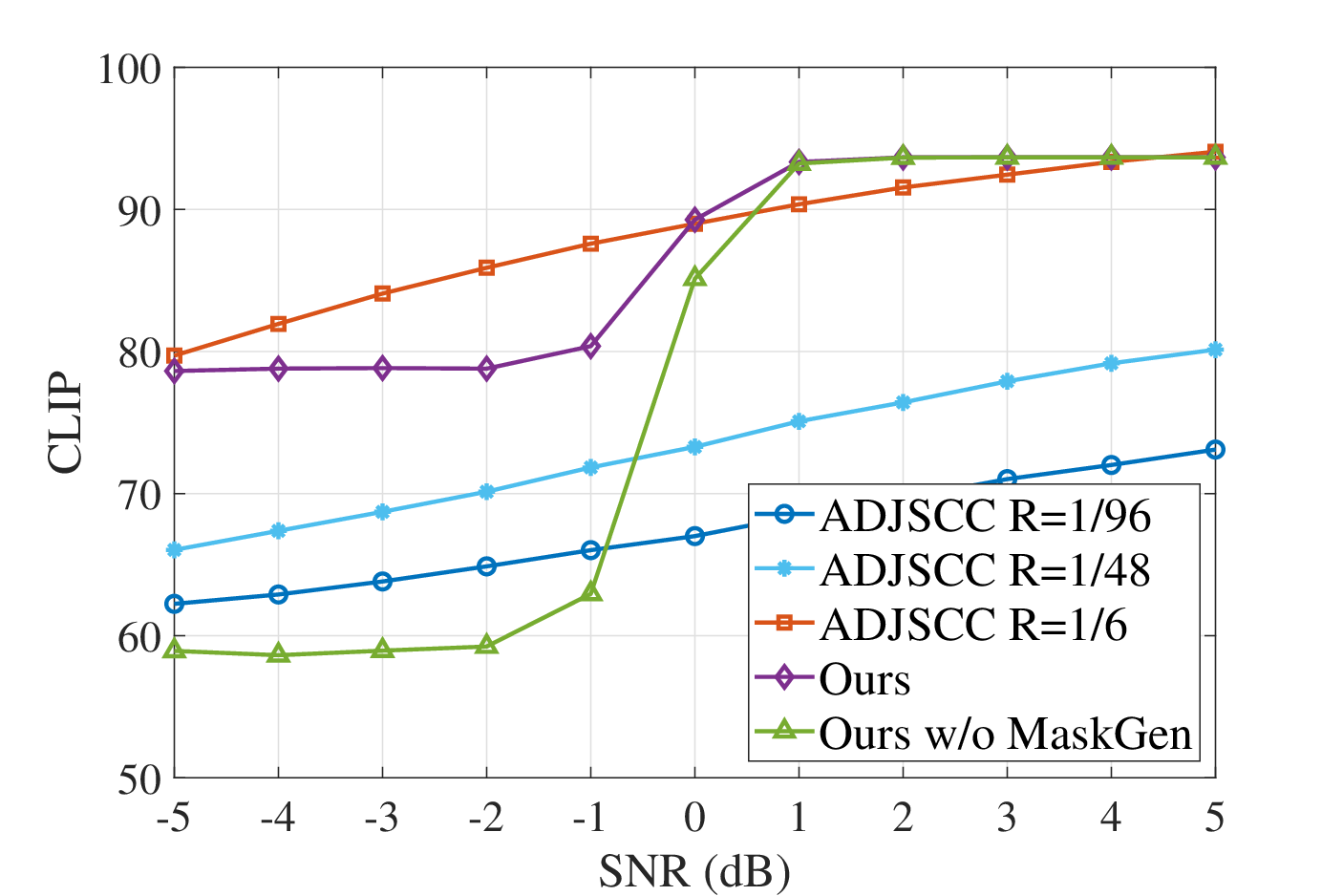}}
    \caption{\small Performance comparison between our proposed token-based system and ADJSCC at different bandwidth bandwidth ratios on the ImageNet dataset across varying SNR levels. \textbf{(a)} PSNR comparison shows our approach achieving superior pixel-level fidelity at higher SNRs. \textbf{(b)} LPIPS comparison demonstrates better perceptual quality with our method at most SNR levels. \textbf{(c)} CLIP similarity comparison reveals our system's superior semantic preservation capability, particularly important for communication under extreme conditions.}
    \label{fig:three_metrics}
    \vspace{-14pt}
\end{figure*}

\begin{figure*}
    \centering
    \subfigure[PSNR(↑) vs. SNR]{
    \includegraphics[width=0.32\textwidth, trim=8mm 0mm 10mm 0mm, clip]{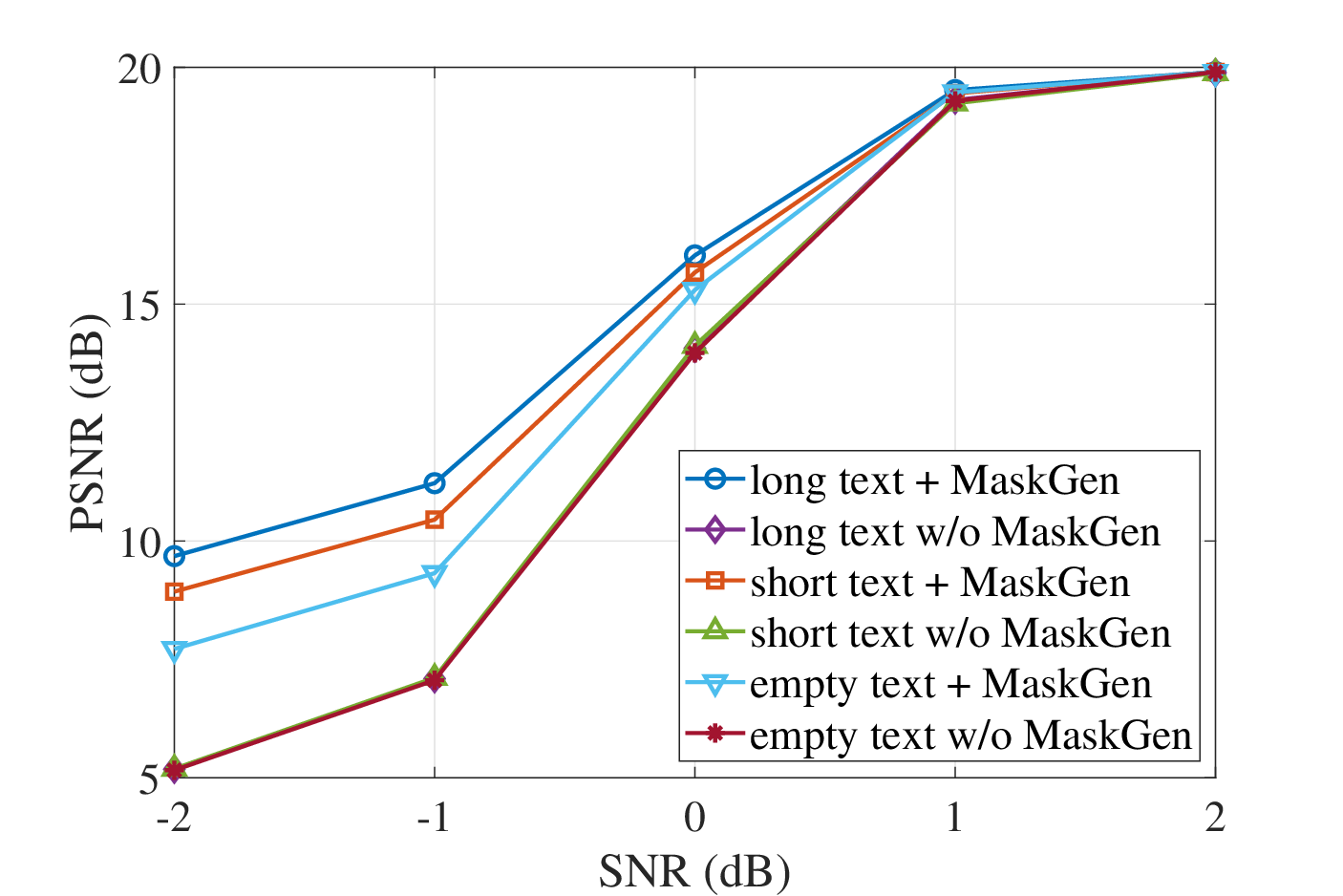}}
    \subfigure[LPIPS(↓) vs. SNR]{
    \includegraphics[width=0.32\textwidth, trim=8mm 0mm 10mm 0mm, clip]{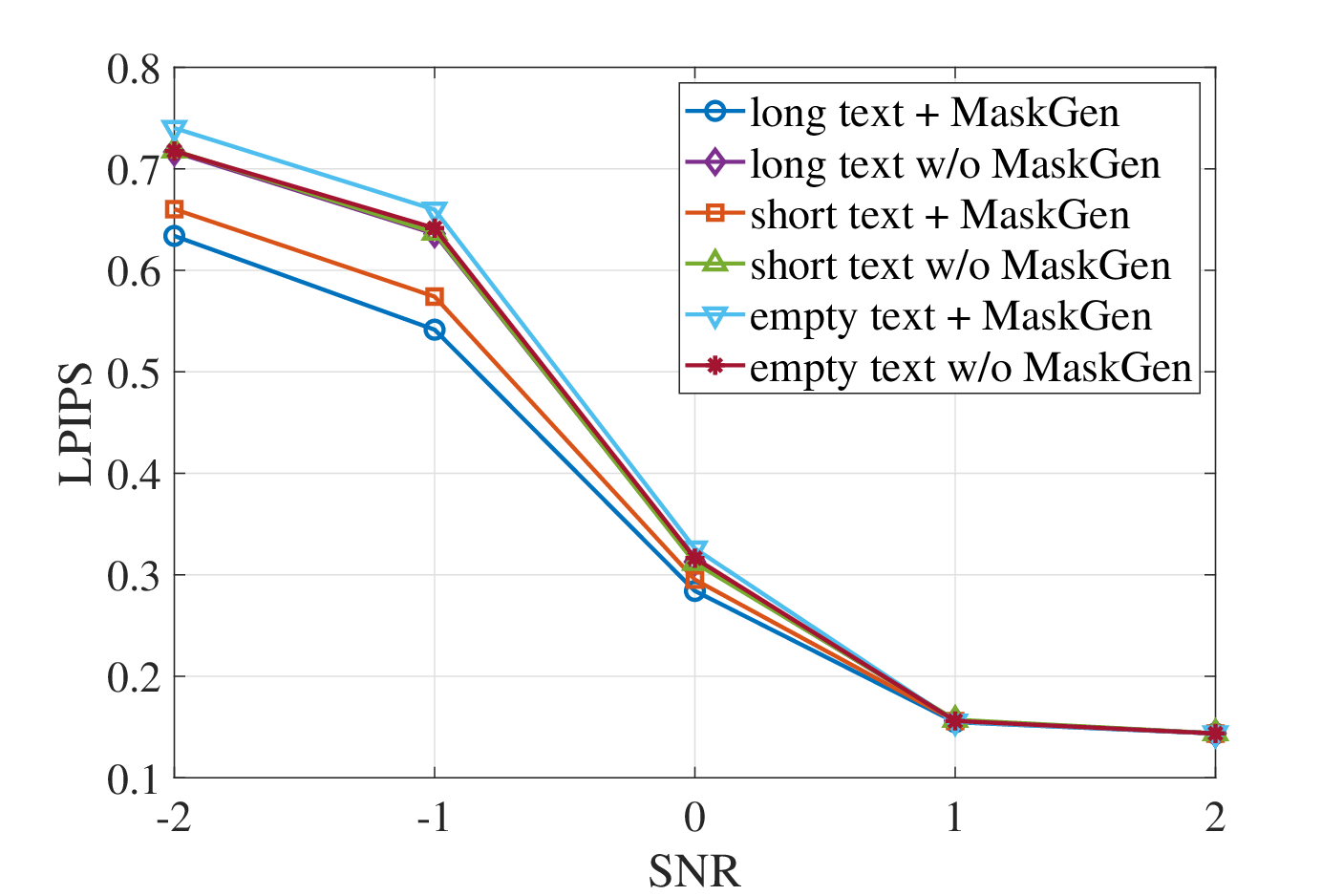}}
    \subfigure[CLIP(↑) vs. SNR]{
    \includegraphics[width=0.32\textwidth, trim=8mm 0mm 10mm 0mm, clip]{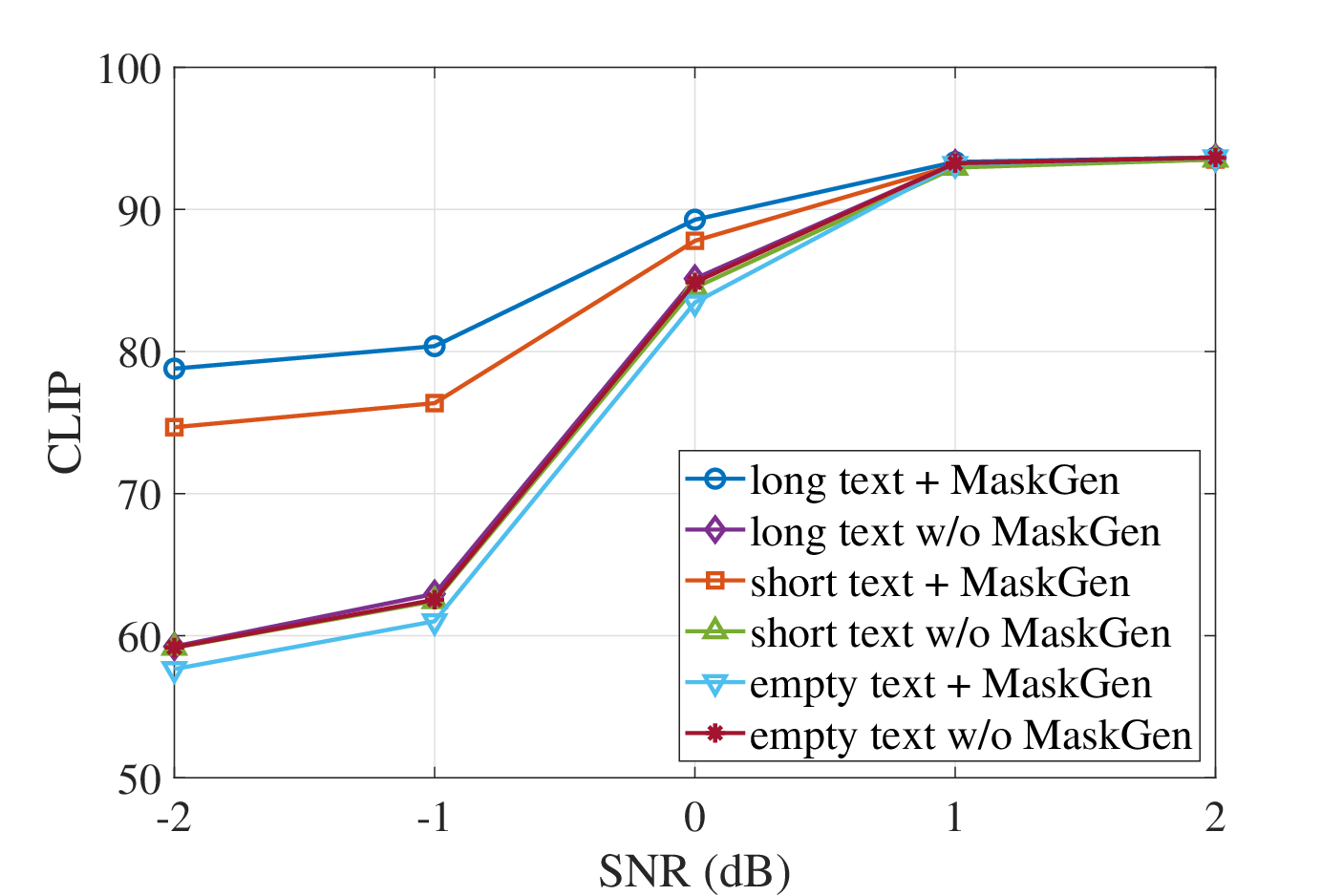}}
    \caption{\small Effect of caption length on reconstruction quality under various SNRs. All metrics (PSNR, LPIPS, and CLIP similarity) consistently demonstrate that longer, more informative captions enhance reconstruction performance, with benefits most pronounced at low SNR conditions and when utilizing the MaskGen module. These results confirm that detailed captions provide valuable guidance for token prediction during challenging image reconstruction scenarios.}
    \label{fig:text_length}
    \vspace{-14pt}
\end{figure*}

\section{Experimental Results and Analysis}

\subsection{Experimental Setup}
We conducted all transmission experiments on a single NVIDIA RTX 2080Ti GPU. The polar code implementation utilized the NVIDIA Sionna library \cite{sionna} for efficient channel coding and decoding operations. The TA-TiTok tokenizer \cite{kim2025demo} employs a codebook of size 8192, with each image represented by 128 tokens. For channel coding, we process 8 tokens as a single package, resulting in 104 information bits per package (13 bits per token) and a code length of 256 bits. We employ 4-QAM constellation modulation, achieving an overall bandwidth bandwidth ratio of R = 1/96. 
Our evaluations focus on AWGN channels with SNRs from -5 dB to 5 dB, representing challenging communication conditions relevant to edge deployment scenarios.

\subsubsection{Dataset Preparation}
Following preprocessing approaches used in pre-trained foundation models, we selected images from ImageNet \cite{imagenet} with aspect ratios less than 2 and longer sides larger than 256 pixels. We applied additional filtering using LAION watermark detection \cite{laion_water} and aesthetic prediction \cite{laion_aes} to retain only non-watermarked samples with aesthetic scores exceeding 6. For text descriptions, we employed Molmo-7B \cite{molmo} to generate accurate captions for each image within the CLIP encoder's 77-token limit.

To ensure a fair evaluation, we conducted our experiments exclusively on datasets that were not encountered by the pre-trained model during its training phase.
\subsubsection{Evaluation Metrics}
We evaluated our system using multiple metrics to assess both perceptual and statistical aspects of reconstruction quality:
\begin{itemize}
    \item PSNR (Peak Signal-to-Noise Ratio): Measures pixel-level fidelity between original and reconstructed images.
    
    \item LPIPS (Learned Perceptual Image Patch Similarity) \cite{zhang2018perceptual}: Evaluates perceptual quality based on features from pre-trained networks.
    
    \item CLIP similarity \cite{clip}\cite{clip-score}: Measures semantic preservation by computing cosine similarity between CLIP embeddings.
\end{itemize}

This multi-metric evaluation comprehensively assesses our system's performance across different dimensions of image quality relevant to human perception.

\subsection{Quantitative Results}
We compared our token-based system with deep source channel coding with attention modules (ADJSCC) \cite{xu2021wireless} method at bandwidth bandwidth ratios of R=1/6, R=1/48, and R=1/96, with our system operating at R=1/96. For fair comparison with our text-enhanced approach, we primarily contrast against ADJSCC at R=1/48, as we use additional text modality.

As shown in Fig.~\ref{fig:three_metrics}, our system demonstrates superior perceptual quality (LPIPS) and semantic preservation (CLIP similarity) at SNR values above 0 dB, despite lower performance in pixel-level fidelity (PSNR). The text-guided MaskGen model enhances performance and mitigates the cliff effect at lower SNRs - a phenomenon where reconstruction quality deteriorates rapidly rather than gracefully when the SNR falls below a critical threshold. Higher CLIP similarity scores confirm our approach's effectiveness in preserving semantic content, which aligns with prioritizing perceptually important features.

\begin{figure*}[t]
    \centering
    \setlength{\tabcolsep}{1pt}
    \begin{tabular}{p{1.5cm}ccccc} \small\textbf{Raw} &
    \includegraphics[width=0.14\textwidth]{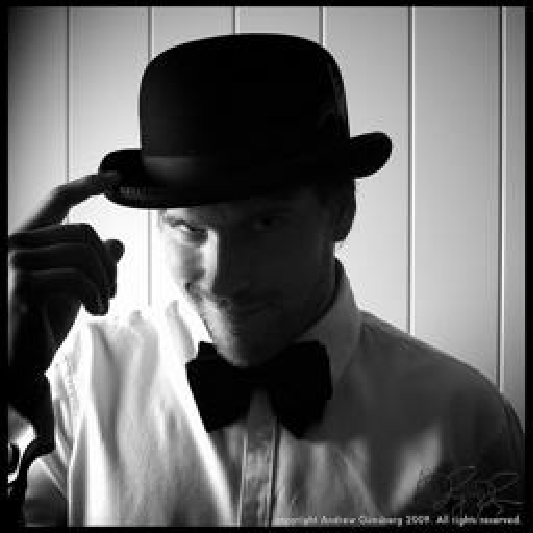} &
    \includegraphics[width=0.14\textwidth]{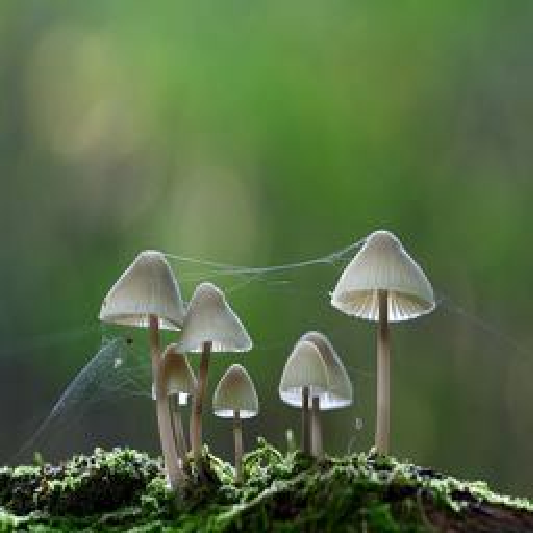} &
    \includegraphics[width=0.14\textwidth]{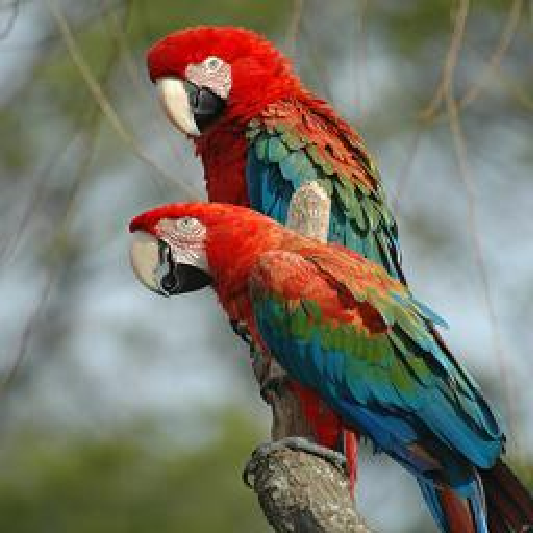} &
    \includegraphics[width=0.14\textwidth]{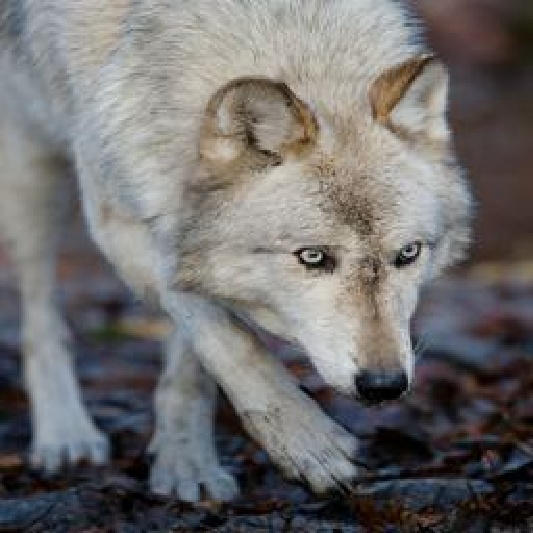} &
    \includegraphics[width=0.14\textwidth]{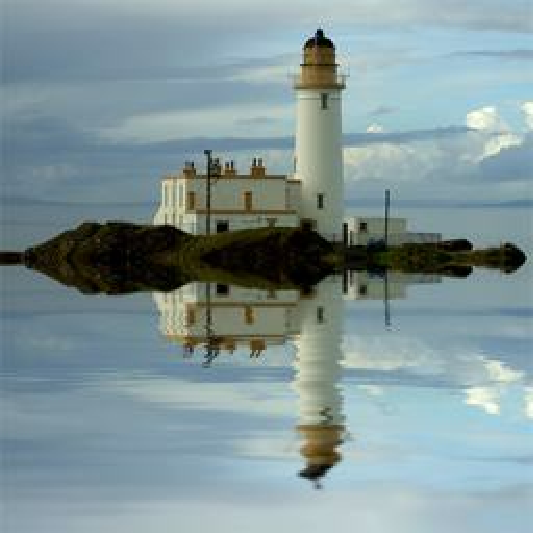} \\[0em]
    
    \small\textbf{Ours} &
    \includegraphics[width=0.14\textwidth]{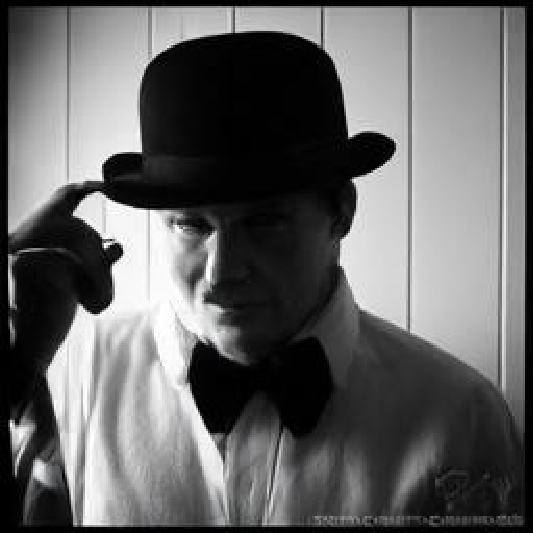} &
    \includegraphics[width=0.14\textwidth]{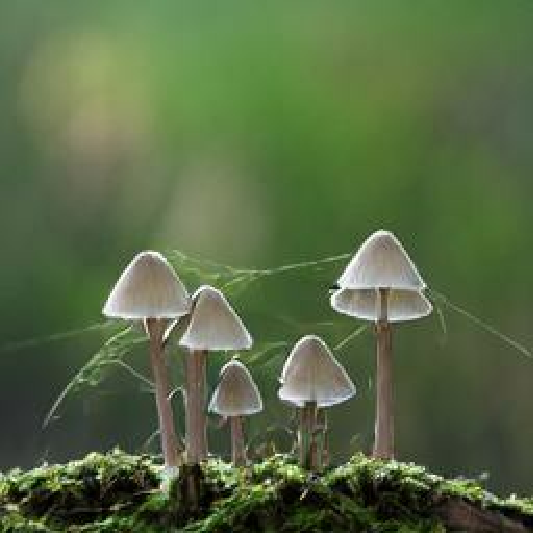} &
    \includegraphics[width=0.14\textwidth]{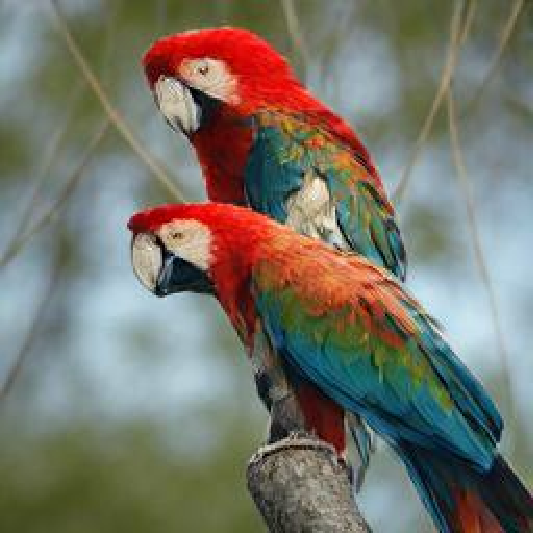} &
    \includegraphics[width=0.14\textwidth]{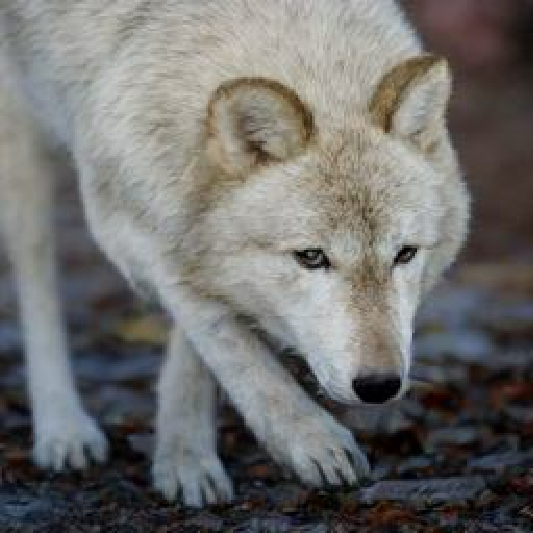} &
    \includegraphics[width=0.14\textwidth]{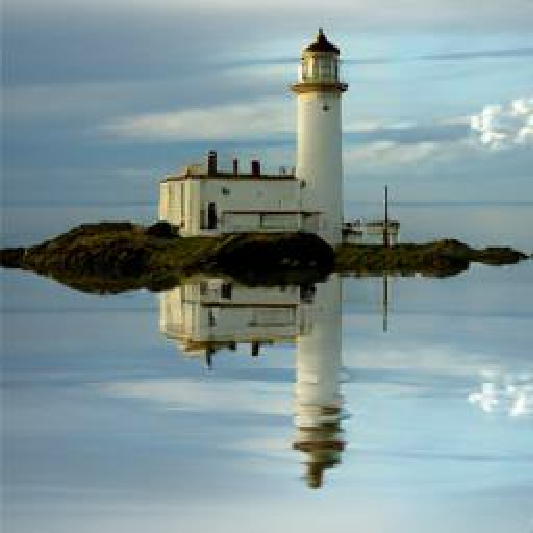} \\[0em]
    
    \small\textbf{ADJSCC} &
    \includegraphics[width=0.14\textwidth]{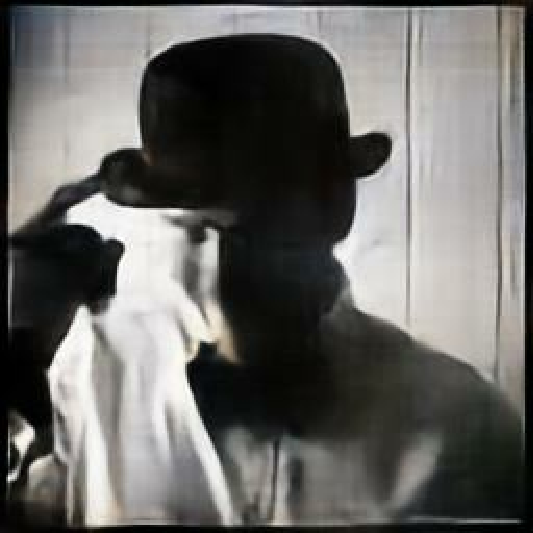} &
    \includegraphics[width=0.14\textwidth]{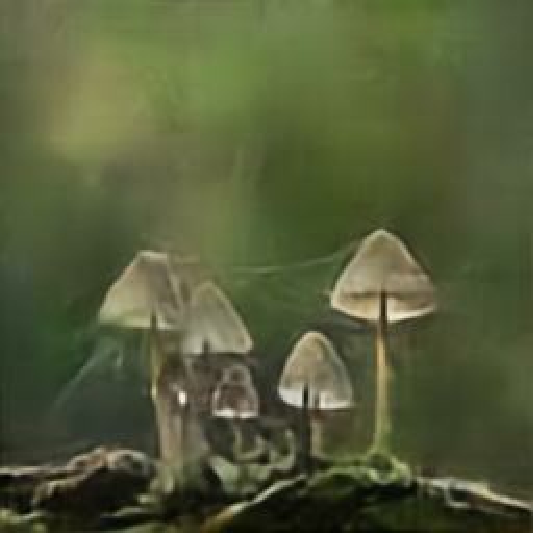} &
    \includegraphics[width=0.14\textwidth]{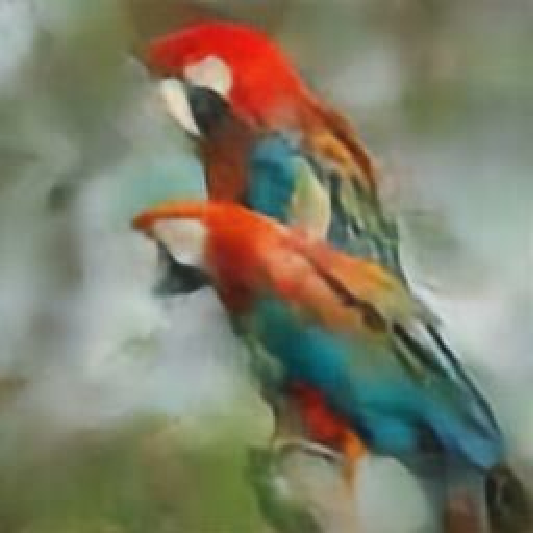} &
    \includegraphics[width=0.14\textwidth]{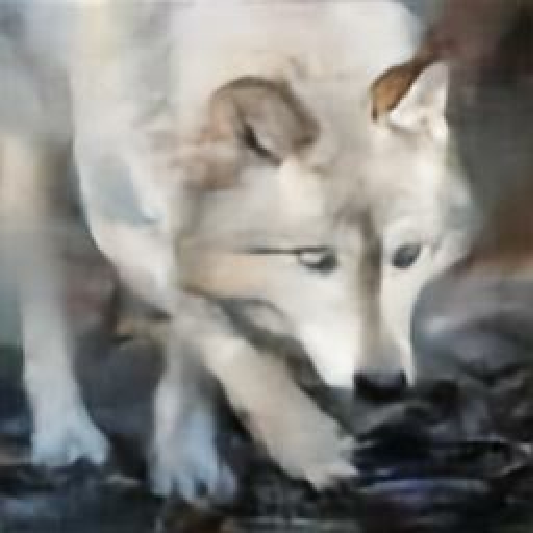} &
    \includegraphics[width=0.14\textwidth]{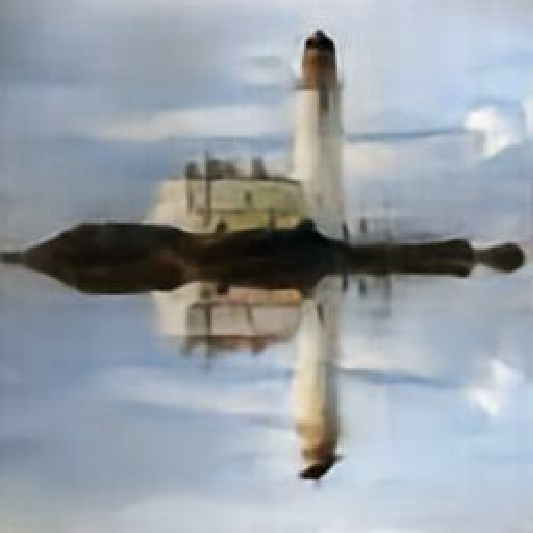} \\
    \end{tabular}
    \caption{Image recovery comparison at SNR=1dB. Our method preserves better semantic content compared to ADJSCC R=1/48.}
    \label{fig:recovery_comparison}
    \vspace{-14pt}
\end{figure*}

 Fig.~\ref{fig:recovery_comparison} presents a visual comparison between our approach and the ADJSCC (R=1/48) method for image reconstruction at SNR=1dB. As clearly demonstrated, our method preserves more semantic details and produces visually superior results with better structural integrity ,particularly in complex regions of the images. The enhanced visual quality aligns with our quantitative metrics, confirming the effectiveness of our token-based transmission system.

\subsection{Ablation Studies}
We investigated the impact of textual description length on ImageNet (Fig.~\ref{fig:text_length}). Our experiments compared long captions ($\sim$70 tokens), short captions ($\sim$30 tokens), and empty captions as baseline. The results demonstrate that longer and more detailed captions significantly enhance reconstruction quality at low SNR levels when utilizing the MaskGen module. However, these benefits show diminishing returns as SNR levels increase. This finding highlights the considerable potential of token prediction mechanisms in semantic image transmission systems. Notably, our text-guided approach substantially outperforms method that do not leverage textual information in low SNR conditions with the MaskGen module, thus demonstrating the inherent cross-modality resilience of our proposed framework.

\begin{figure}
    \centering
    \includegraphics[width=0.45\textwidth, trim=10mm 0mm 10mm 5mm, clip]{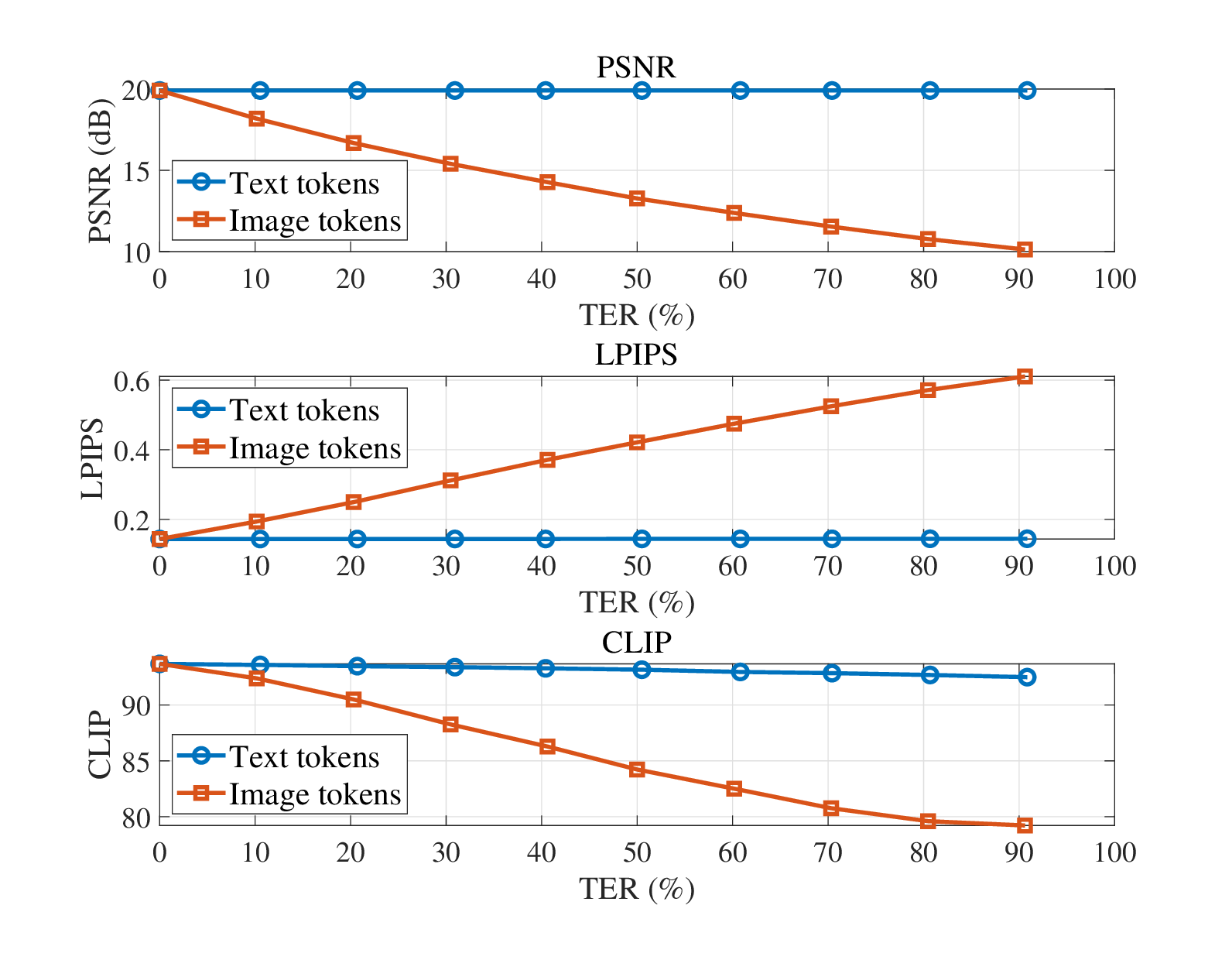}
    \vspace{-12pt}
    \caption{\small Performance comparison under different modality error conditions. Image token integrity is more critical for reconstruction, while text tokens provide complementary support in semantic perception.}
    \label{fig:multimodal}
    \vspace{-12pt}
\end{figure}

To understand the interplay between modalities, we varied error rates in both simultaneously (Fig.~\ref{fig:multimodal}). Our analysis reveals an asymmetric contribution pattern: when image tokens are intact, text errors only have marginal impact on semantic perception metrics; conversely, reconstruction quality deteriorates significantly as image token corruption increases, demonstrating the dominant role of visual information in multimodal representations.

This asymmetry suggests that while textual information provides supplementary semantic context, the fidelity of the underlying visual tokens remains the primary determinant of overall system performance in cross-modal transmission scenarios. Such findings indicate potential benefits from optimizing bit allocation between modalities based on their relative importance—a promising direction for future work in adaptive multi-modal resource allocation for bandwidth-constrained communication systems. 

\begin{figure*}[]
    \centering
    \subfigure[LPIPS(↓) vs. SNR]{
    \includegraphics[width=0.45\textwidth, trim=10mm 0mm 10mm 0mm, clip]{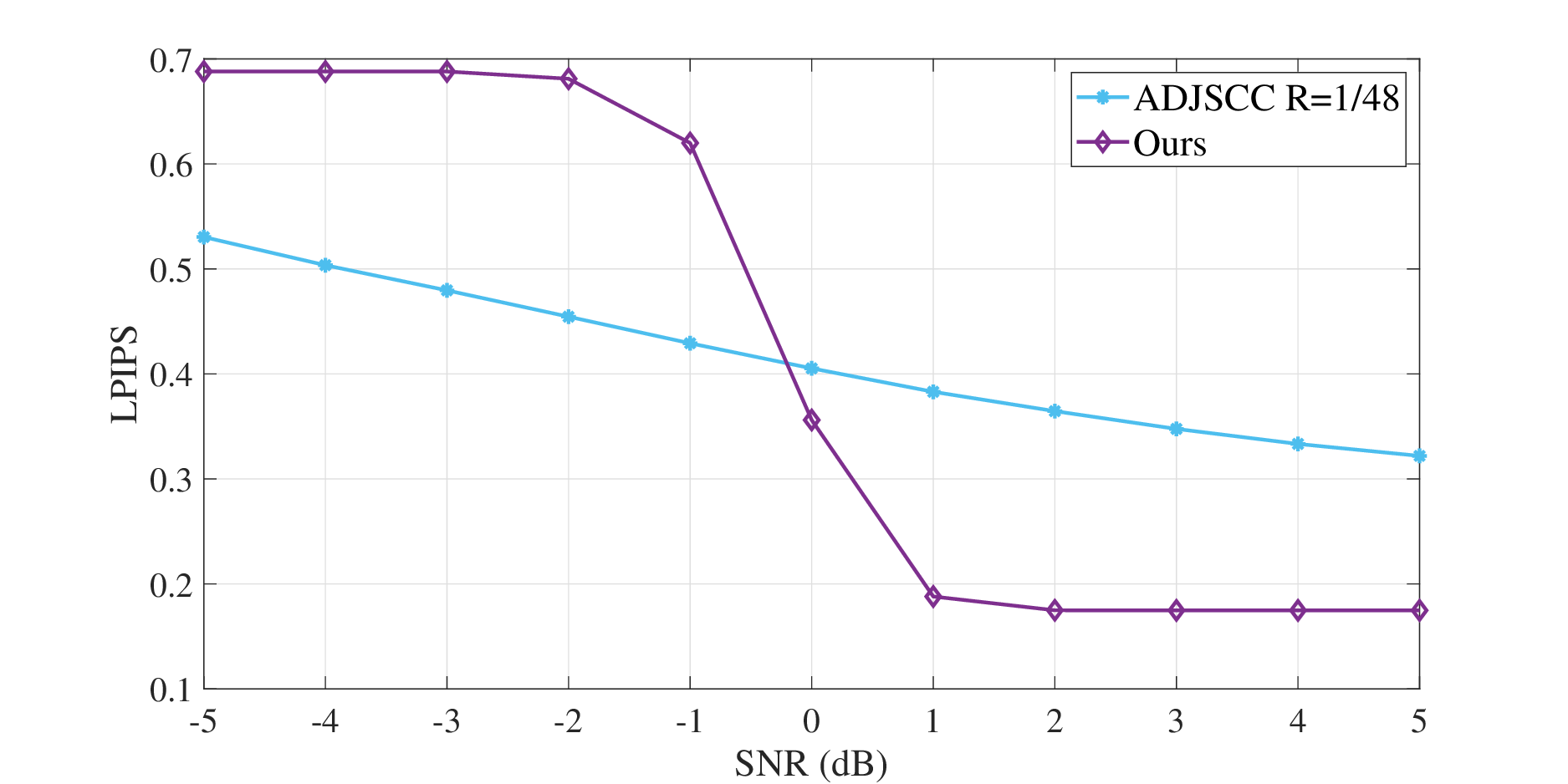}}
    \subfigure[CLIP(↑) vs. SNR]{
    \includegraphics[width=0.45\textwidth, trim=10mm 0mm 10mm 0mm, clip]{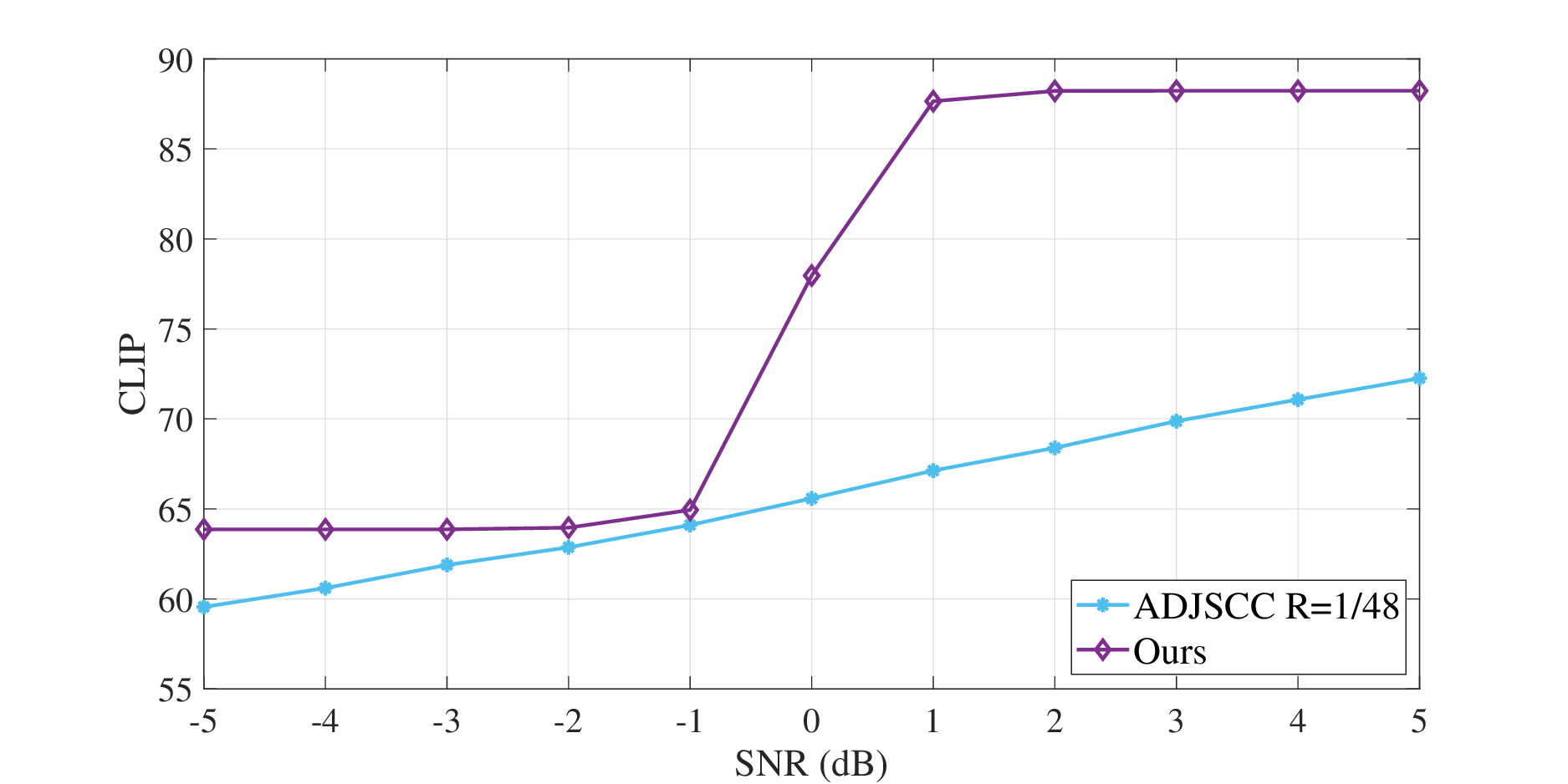}}
    \caption{\small Generalization performance comparison on the flickr dataset. \textbf{(a)} LPIPS scores comparing our approach with ADJSCC (R=1/48) across SNR levels. \textbf{(b)} CLIP similarity comparison showing superior semantic preservation with our method. Both models were not specifically trained on flickr, demonstrating our system's superior generalization capability to unseen data distributions.}
    \label{fig:flickr}
    \vspace{-14pt}
\end{figure*}

Additionally, we evaluated the generalization capability of our approach on the Flickr\cite{flickr} dataset (Fig.~\ref{fig:flickr}). Our token-based method demonstrates superior performance compared to ADJSCC (R=1/48), achieving lower LPIPS values when SNR exceeds 0 dB, while consistently maintaining higher CLIP similarity scores across all SNR ranges. Notably, our approach achieves an LPIPS score as low as 0.17, demonstrating exceptional perceptual quality reconstruction. These results highlight that our method inherits the robust representations of its underlying pre-trained models—eliminating the need for costly domain-specific retraining.

\subsection{Computational Efficiency Analysis}
\begin{table}
    \centering
    \caption{Computational resource comparison between our token-based approach and ADJSCC.}
    \begin{tabular}{|c|c|c|}
        \hline
        \textbf{Model} & \textbf{Inference FLOPs\cite{ptflops}} & \textbf{Retraining cost} \\
        \hline
        ADJSCC (R=1/48) & 92.4G & $\sim$10 GPU hours\\
        Ours (w/o MaskGen) & 358.8G & 0  \\
        One step MaskGen & 325.3G & 0  \\
        \hline
    \end{tabular}
    \label{tab:computation}
    \vspace{-14pt}
\end{table}

Table \ref{tab:computation} presents a comprehensive comparison of computational resources required by our token-based approach versus ADJSCC across the entire transmission pipeline, including both tokenization and detokenization processes. The metric ``One step MaskGen'' represents a single iteration of our token prediction mechanism using the MaskGen module. While our foundation model based system has higher inference floating point operations (FLOPs) due to the larger model size, it completely eliminates training overhead when facing a new channel conditions—only adjustment of channel coding parameters is needed. 

These complementary characteristics—cross-dataset generalization and channel adaptability—reduce deployment complexity in practical wireless scenarios, despite higher computational requirements per inference. By eliminating retraining needs for new visual domains and varying channel qualities, our approach enables efficient deployment across diverse network environments. While acknowledging current computational trade-offs, our foundation model-based approach represents a shift toward more robust semantic communication systems aligned with next-generation content delivery requirements.
\vspace{-5pt}
\section{Conclusion}
We presented a novel text-guided token communication system that leverages pre-trained generative foundation models for wireless image transmission. Our system integrates textual guidance with image tokens to maintain robust performance under challenging channel conditions. Simulation results confirm that our approach outperforms the compared ADJSCC method in perceptual quality (LPIPS) and semantic preservation (CLIP similarity) at SNR levels above 0 dB with bandwidth ratio of 1/96, while mitigating the cliff effect at lower SNRs through text-guided token predictions.

The key advantages of our approach are twofold: (1) enhanced cross-modal resilience through text guidance that complements degraded visual information, as demonstrated in our ablation studies; and (2) 
generalization capability that eliminates domain-specific retraining costs, allowing immediate adaptation to unseen scenarios with only parameter adjustments rather than the approximately 10 GPU hours required by comparable methods. Future research directions include adaptive modality weighting based on channel conditions, enhanced token prediction mechanisms, and extension to video transmission. 

\vspace{-10pt}
\section*{Acknowledgement}
The work was supported by the Natural Science Foundation of China (NSFC) under Grant 62471036, Shandong Province Natural Science Foundation under Grant ZR2022YQ62, Beijing Natural Science Foundation under Grant L242011, and Beijing Nova Program.

\bibliographystyle{ieeetr}
\bibliography{references}

\begin{thebibliography}{10}

\bibitem{chen20236g}
C.~Chen, C.~Wang, C.~Li, {\em et~al.}, ``A v2v emergent message dissemination scheme for {6G}-oriented vehicular networks,'' {\em Chinese Journal of Electronics}, 2023.

\bibitem{gao20246g}
Z.~Gao, D.~Mi, C.~Jiang, {\em et~al.}, ``Emerging space communication and network technologies for {6G} ubiquitous connectivity,'' {\em Space Sci. Technol.}, 2024.

\bibitem{Wang20236g}
Y.~Wang, Z.~Gao, D.~Zheng, {\em et~al.}, ``Transformer-empowered {6G} intelligent networks: From massive {MIMO} processing to semantic communication,'' {\em {IEEE} Wireless Commun.}, vol.~30, pp.~127--135, 2022.

\bibitem{Liu2024space}
H.~Liu, T.~Qin, Z.~Gao, {\em et~al.}, ``Near-space communications: The last piece of {6G} space-air-ground-sea integrated network puzzle,'' {\em Space Sci. Technol.}, vol.~4, p.~0176, Aug. 2024.

\bibitem{bourtsoulatze2019deep}
E.~Bourtsoulatze, D.~B. Kurka, and D.~G{\"u}nd{\"u}z, ``Deep joint source-channel coding for wireless image transmission,'' {\em Proc. IEEE Int. Conf. Acoust. Speech Sig. Process.}, pp.~4774--4778, 2018.

\bibitem{xu2021wireless}
J.~lin Xu, B.~Ai, W.~Chen, {\em et~al.}, ``Wireless image transmission using deep source channel coding with attention modules,'' {\em {IEEE} Trans. Circuits Syst. Video Technol.}, vol.~32, pp.~2315--2328, 2020.

\bibitem{wu2025semantic}
M.~Wu, Z.~Gao, Z.~Wang, {\em et~al.}, ``Deep joint semantic coding and beamforming for near-space airship-borne massive {MIMO} network,'' {\em {IEEE} J. Select. Areas Commun.}, vol.~43, pp.~260--278, Jan. 2025.

\bibitem{shi2021semantic}
T.-Y. Tung, D.~B. Kurka, M.~Jankowski, {\em et~al.}, ``Deepjscc-q: Constellation constrained deep joint source-channel coding,'' {\em {IEEE} J. Sel. Areas Inf. Theory}, vol.~3, pp.~720--731, 2022.

\bibitem{qiao2024wcl}
L.~Qiao, M.~B. Mashhadi, Z.~Gao, {\em et~al.}, ``Latency-aware generative semantic communications with pre-trained diffusion models,'' {\em {IEEE} Wireless Commun. Lett.}, vol.~13, pp.~2652--2656, 2024.

\bibitem{gao2023semantic}
Z.~Gao, S.~Liu, Y.~Su, {\em et~al.}, ``Hybrid knowledge-data driven channel semantic acquisition and beamforming for cell-free massive {MIMO},'' {\em {IEEE} J. Sel. Topics Signal Process.}, vol.~17, pp.~964--979, 2023.

\bibitem{van2017neural}
A.~van~den Oord, O.~Vinyals, and K.~Kavukcuoglu, ``Neural discrete representation learning,'' in {\em Proc. Adv. Neural Inf. Process. Syst. (NeurIPS)}, NIPS'17, p.~6309–6318, Curran Associates Inc., 2017.

\bibitem{esser2021taming}
P.~Esser, R.~Rombach, and B.~Ommer, ``Taming transformers for high-resolution image synthesis,'' in {\em Proc. IEEE Conf. Comput. Vis. Pattern Recognit. (CVPR)}, pp.~12873--12883, June 2021.

\bibitem{kim2025demo}
D.~Kim, J.~He, Q.~Yu, {\em et~al.}, ``Democratizing text-to-image masked generative models with compact text-aware one-dimensional tokens,'' {\em arXiv preprint arXiv:2501.07730}, 2025.

\bibitem{muse}
H.~Chang, H.~Zhang, J.~Barber, {\em et~al.}, ``Muse: Text-to-image generation via masked generative transformers,'' in {\em Proc. Int. Conf. Mach. Learn. (ICML)}, vol.~202 of {\em Proceedings of Machine Learning Research}, pp.~4055--4075, PMLR, July 2023.

\bibitem{qiao2025tokcom}
L.~Qiao, M.~Boloursaz~Mashhadi, Z.~Gao, {\em et~al.}, ``Token communications: A large model-driven framework for cross-modal context-aware semantic communications,'' {\em arXiv preprint arXiv:2502.12096}, 2025.

\bibitem{lee2025semantic}
S.~Lee, J.~Park, J.~Choi, {\em et~al.}, ``Semantic packet aggregation for token communication via genetic beam search,'' {\em arXiv preprint arXiv:2504.19591}, 2025.

\bibitem{qiao2025todma}
L.~Qiao, M.~Boloursaz~Mashhadi, Z.~Gao, and D.~G{\"u}nd{\"u}z, ``Token-domain multiple access: Exploiting semantic orthogonality for collision mitigation,'' {\em arXiv preprint arXiv:2502.06118}, 2025.

\bibitem{qiao2025todma_large}
L.~Qiao, M.~B. Mashhadi, Z.~Gao, R.~Schober, and D.~G{\"u}nd{\"u}z, ``{ToDMA}: Large model-driven token-domain multiple access for semantic communications,'' {\em arXiv preprint arXiv:2505.10946}, May 2025.

\bibitem{3gpp38212}
3GPP, ``{5G}; {NR}; multiplexing and channel coding,'' Technical Specification (TS) 38.212, {3rd Generation Partnership Project (3GPP)}, Mar. 2021.
\newblock Version 16.5.0.

\bibitem{sionna}
J.~Hoydis, S.~Cammerer, F.~{Ait Aoudia}, {\em et~al.}, ``Sionna,'' 2022.
\newblock https://nvlabs.github.io/sionna/.

\bibitem{clip}
A.~Radford, J.~W. Kim, C.~Hallacy, {\em et~al.}, ``Learning transferable visual models from natural language supervision,'' in {\em Proc. Int. Conf. Mach. Learn. (ICML)}, vol.~139 of {\em Proceedings of Machine Learning Research}, pp.~8748--8763, PMLR, July 2021.

\bibitem{imagenet}
J.~Deng, W.~Dong, R.~Socher, L.-J. Li, K.~Li, and L.~Fei-Fei, ``{ImageNet}: A large-scale hierarchical image database,'' in {\em Proc. IEEE Conf. Comput. Vis. Pattern Recognit. (CVPR)}, pp.~248--255, 2009.

\bibitem{laion_water}
``{LAION-5B-WatermarkDetection}.'' Available: \url{https://github.com/LAION-AI/LAION-5B-WatermarkDetection}.

\bibitem{laion_aes}
``{LAION2B-en-aesthetic}.'' Available: \url{https://huggingface.co/datasets/laion/laion2B-en-aesthetic}.

\bibitem{molmo}
M.~Deitke, C.~Clark, S.~Lee, {\em et~al.}, ``Molmo and pixmo: Open weights and open data for state-of-the-art vision-language models,'' in {\em Proc. IEEE Conf. Comput. Vis. Pattern Recognit. (CVPR)}, pp.~91--104, June 2025.

\bibitem{zhang2018perceptual}
R.~Zhang, P.~Isola, A.~A. Efros, {\em et~al.}, ``The unreasonable effectiveness of deep features as a perceptual metric,'' in {\em Proc. IEEE Conf. Comput. Vis. Pattern Recognit. (CVPR)}, June 2018.

\bibitem{clip-score}
J.~Hessel, A.~Holtzman, M.~Forbes, {\em et~al.}, ``{CLIPS}core: A reference-free evaluation metric for image captioning,'' in {\em Proceedings of the 2021 Conference on Empirical Methods in Natural Language Processing}, pp.~7514--7528, Association for Computational Linguistics, Nov. 2021.

\bibitem{flickr}
P.~Young, A.~Lai, M.~Hodosh, {\em et~al.}, ``From image descriptions to visual denotations: New similarity metrics for semantic inference over event descriptions,'' {\em Transactions of the Association for Computational Linguistics}, vol.~2, pp.~67--78, 2014.

\bibitem{ptflops}
V.~Sovrasov, ``ptflops: a flops counting tool for neural networks in pytorch framework.'' Available: \url{https://github.com/sovrasov/flops-counter.pytorch}, 2018-2024.

\end{thebibliography}

\end{document}